\documentclass[
10pt,
showpacs,preprintnumbers,nofootinbib,
 amsmath,amssymb,
 aps,
twocolumn,groupedaddress,superscriptaddress,
floatfix,
showkeys
]{revtex4-1}

\usepackage[labelfont={small},subrefformat=parens,caption=false]{subfig}
\usepackage[dvipsnames]{xcolor}
\usepackage{amsfonts,amsmath,amssymb,bm}
\usepackage{graphicx}
\usepackage[utf8]{inputenc}
\usepackage{xspace}
\usepackage{isotope}
\usepackage{xparse}

\usepackage[pdfpagelabels, pdfencoding=auto, psdextra]{hyperref}
\hypersetup{%
 pdfsubject=Paper,
 pdfkeywords={nuclear physics} {Bayesian} {chiral EFT} {scattering},
 unicode = true,
 breaklinks = true,
 colorlinks = true,
 linkcolor = blue,
 citecolor = blue,
 menucolor = blue,
 citecolor = blue,
 urlcolor = blue
}

\usepackage{silence}
\makeatletter
\newcommand\newsubcommand[3]{\newcommand#1{#2\sc@sub{#3}}}
\def\sc@sub#1{\def\sc@thesub{#1}\@ifnextchar_{\sc@mergesubs}{_{\sc@thesub}}}
\def\sc@mergesubs_#1{_{\sc@thesub#1}}

\newcommand\newsupcommand[3]{\newcommand#1{#2\sc@sup{#3}}}
\def\sc@sup#1{\def\sc@thesup{#1}\@ifnextchar^{\sc@mergesups}{^{\sc@thesup}}}
\def\sc@mergesups^#1{^{\sc@thesup#1}}
\makeatother

\DeclareMathAlphabet{\mathbcal}{OMS}{cmsy}{b}{n}

\newcommand{\boldvec}[1]{\bm{#1}}

\newcommand{\ordervec}{\vec}

\newcommand{\inputvec}{\mathbf}

\newcommand{\rvec}{\boldvec{r}}

\newsubcommand{\ckvec}{\ordervec{c}}{k}

\newsubcommand{\bkvec}{\ordervec{b}}{k}

\newsubcommand{\ckvecset}{\ordervec{\inputvec{c}}}{k}

\newsubcommand{\ckvecapprox}{\mathbf{c}'}{k}
\newsubcommand{\ckvecapproxset}{\mathbf{C}'}{k}

\newsubcommand{\bkvecapprox}{\mathbf{b}'}{k}
\newsubcommand{\bkvecset}{\mathbf{B}}{k}
\newsubcommand{\bkvecapproxset}{\mathbf{B}'}{k}

\newcommand{\genobs}{y}

\newsubcommand{\genobsvec}{\ordervec{\genobs}}{k}
\newsubcommand{\genobsvecset}{\ordervec{\inputvec{\genobs}}}{k}

\newsubcommand{\akvec}{\mathbf{a}}{k}

\newsubcommand{\akvecapprox}{\mathbf{a}'}{k}
\newsubcommand{\akvecset}{\mathbf{A}}{k}
\newsubcommand{\akvecapproxset}{\mathbf{A}'}{k}

{}  

\newcommand{\thetavec}{\bm{\theta}}

\def\diffd{\mathrm{d}}  

\DeclareDocumentCommand\differential{ o g d() }{ 
    \IfNoValueTF{#2}{
        \IfNoValueTF{#3}
            {\diffd\IfNoValueTF{#1}{}{^{#1}}}
            {\mathinner{\diffd\IfNoValueTF{#1}{}{^{#1}}\argopen(#3\argclose)}}
        }
        {\mathinner{\diffd\IfNoValueTF{#1}{}{^{#1}}#2} \IfNoValueTF{#3}{}{(#3)}}
    }

\newcommand{\pathd}{\mathcal{D}}  

\DeclareDocumentCommand\pathdifferential{ o g d() }{ 
    \IfNoValueTF{#2}{
        \IfNoValueTF{#3}
            {\pathd\IfNoValueTF{#1}{}{^{#1}}}
            {\mathinner{\pathd\IfNoValueTF{#1}{}{^{#1}}\argopen(#3\argclose)}}
        }
        {\mathinner{\pathd\IfNoValueTF{#1}{}{^{#1}}#2} \IfNoValueTF{#3}{}{(#3)}}
    }

\newcommand{\tauex}{\tau_{\text{exact}}}

\newcommand{\Hhat}{\widehat H}

\newcommand{\That}{\widehat T}
\newcommand{\Vhat}{\widehat V}
\newcommand{\psigs}{\psi_{\text{gs}}}
\newcommand{\utrial}{u_{\text{t}}}
\newcommand{\uexact}{u_{\text{exact}}}

\newcommand{\singletS}{$^1$S$_0$}
\newcommand{\tripletS}{$^3$S$_1$}

\newcommand{\psit}{\psi_{\text{trial}}}
\newcommand{\psitbra}{\langle\psit |}
\newcommand{\psitket}{|\psit\rangle}
\newcommand{\Utilde}{\widetilde U}

\newcommand{\angL}{\ell}

\newcommand{\Km}{\mathcal{K}}
\newcommand{\Nb}{N_b} 
\newcommand{\EVC}{EC} 

\usepackage{amsmath,environ}
\NewEnviron{EqS}{%
\begin{equation}\begin{split}
  \BODY
\end{split}\end{equation}
}
\NewEnviron{EqSn}{%
\begin{equation*}\begin{split}
  \BODY
\end{split}\end{equation*}
}

\begin{document}

\title{Efficient emulators for scattering using eigenvector continuation}

\author{R.~J. Furnstahl}
\email{furnstahl.1@osu.edu}
\affiliation{Department of Physics, The Ohio State University, Columbus, OH
43210, USA}

\author{A.~J. Garcia}
\email{garcia.823@osu.edu}
\affiliation{Department of Physics, The Ohio State University, Columbus, OH
43210, USA}

\author{P.~J. Millican}
\email{millican.7@osu.edu}
\affiliation{Department of Physics, The Ohio State University, Columbus, OH
43210, USA}

\author{Xilin Zhang}
\email{zhang.10038@osu.edu}
\affiliation{Department of Physics, The Ohio State University, Columbus, OH 43210, USA}

\date{\today}

\begin{abstract}
Eigenvector continuation (\EVC) has been shown to accurately and efficiently reproduce ground states for targeted sets  of Hamiltonian parameters. It uses as variational basis vectors the corresponding ground-state eigensolutions from selected other sets of parameters.
Here we extend the \EVC\  approach to scattering using the Kohn variational principle.
We first test it using a model for S-wave nucleon-nucleon scattering and then demonstrate that it also works to give accurate predictions for non-local potentials, charged-particle scattering, complex optical potentials, and higher partial waves.
These proofs-of-principle validate \EVC\  as an accurate emulator for applying Bayesian inference to parameter estimation constrained by scattering observables.
 The efficiency of such emulators is because the accuracy is achieved with a small number of variational basis elements and the central computations are just linear algebra calculations in the space spanned by this basis. 
\end{abstract}

\maketitle

\section{Overview}

Bayesian inference is increasingly favored for uncertainty quantification in nuclear physics calculations (e.g., see~\cite{Furnstahl:2014xsa, Zhang:2015ajn, Wesolowski:2018lzj,Catacora-Rios:2019goa,Neufcourt:2019sle}), but the computational requirements can be substantial.
In particular, Bayesian parameter estimation generally requires Monte Carlo sampling of the parameter space, with many evaluations of the likelihood  with different parameters. 
Each evaluation may be sufficiently expensive that a full parameter estimation is infeasible.
Eigenvector continuation (\EVC)~\cite{Frame:2017fah,Frame:2019jsw} has already shown that it can be used as an efficient and accurate emulator~\cite{Konig:2019adq} to ameliorate this problem.
In applying an emulator, one trains computer models of the relevant calculations using a representative set of parameters and then samples for other parameters from the model instead of full calculations.  
Efficient and effective \EVC\  emulators for nuclear bound-state properties and transitions have been demonstrated for many-body calculations using chiral effective field theory ($\chi$EFT) Hamiltonians~\cite{Konig:2019adq,Ekstrom:2019lss}.

We would also like to have fast \EVC\  emulators for scattering, e.g., for treating reactions and for few-body scattering
used to constrain $\chi$EFT low-energy constants~\cite{Witala:2019ffj}.
The variational method for ground-state energies is well known from elementary quantum mechanics.
In addition, there are variational formulations of scattering, such as those by Schwinger and Kohn (see Refs.~\cite{GOL64,newton2002scattering,taylor2006scattering, nesbet1980variational} and references therein).
The conventional applications in scattering are for two-body scattering, usually in partial waves, but the literature contains adaptations to three-body scattering, including nucleon-deuteron scattering~\cite{Kievsky:1997bg,Kievsky:1997zf}, a process of particular interest for $\chi$EFT~\cite{Witala:2019ffj}.
Here we merge \EVC\  and the Kohn variational principle and explore how well it works using a series of model calculations, starting with two-body scattering in partial waves. As demonstrated below, a small number of variational basis based on EC can reproduce the exact calculations with great accuracy. 
As a result, the main computational cost is just linear algebra in this low-dimensional space.


\section{Formalism}

Consider a Hamiltonian $\Hhat(\thetavec) = \That + \Vhat(\thetavec)$ with adjustable parameters $\thetavec$.
For example, the vector $\thetavec$ could be the depth of a simple square well or the full set of low-energy constants for an effective field theory.
\EVC\  is a variational method that employs a non-orthogonal basis composed of eigenvectors from different parameter sets $\{\thetavec_i\}$ of the Hamiltonian.
For calculating the ground state of $\Hhat(\thetavec)$, the trial wave function is
\begin{align} \label{eq:trialwf}
    \psitket = \sum_{i=1}^{\Nb} c_i |\psigs(\thetavec_i)\rangle
    ,
\end{align}
where $|\psigs(\thetavec_i)\rangle$ is the ground-state eigenvector of $\Hhat(\thetavec_i)$.
(The dependence of $c_i$ and $\psitket$ on $\thetavec_i$ is suppressed for notational convenience.)
The $\Nb$ $\thetavec_i$s are chosen  either systematically or randomly to span a particular range of values, see below.
The effectiveness of the \EVC\  basis can be understood by an analytic continuation analysis~\cite{Frame:2017fah,Sarkar:2020mad}.

The variational principle for the ground-state energy states that the expectation value of $\Hhat(\thetavec)$ in the trial state, subject to the condition that $\psitket$ is normalized, is stationary:
\begin{align} \label{eq:stationarygs}
    \delta \bigl[
      \psitbra \Hhat(\thetavec)\psitket - \lambda (\psitbra \psit \rangle - 1)
    \bigr] = 0 
    .
\end{align}
The stationary solution given Eq.~\eqref{eq:trialwf} is a generalized eigenvalue problem yielding Lagrange multiplier $\lambda_{\text{min}}$, which is an upper bound to $E_{\text{gs}}$, and the $\{c_i\}$ provide an approximation to $|\psigs(\thetavec)\rangle$ through Eq.~\eqref{eq:trialwf}~\cite{Frame:2017fah,Frame:2019jsw,Konig:2019adq,Ekstrom:2019lss}.

For the extension of \EVC\  to scattering we use the Kohn variational principle (KVP)~\cite{Kohn:1948zz,taylor2006scattering}.
There are many variational methods for scattering, but the KVP is particularly straightforward to adapt to \EVC\  in a form similar to Eq.~\eqref{eq:stationarygs}.
Let us start with the goal of finding the phase shift $\delta_\ell(E)$ at energy $E$ for nonrelativistc two-body scattering in an uncoupled partial-wave channel with angular momentum $\ell$ and short-range forces only.
In coordinate space, $\That \rightarrow -\nabla^2/2\mu$ with $\hbar = 1$ and reduced mass $\mu$, and we allow $\Vhat(\thetavec)$ to be local or nonlocal.

We take the trial wave function for the extended \EVC\  to be (we again suppress some $\thetavec_i$ dependence)
\begin{align} \label{eq:trialwfscatt}
    \psitket = \sum_{i=1}^{\Nb} c_i |\psi_E(\thetavec_i)\rangle
    ,
\end{align}
where $|\psi_E(\thetavec_i)\rangle$ is the partial-wave solution for the Schr\"odinger equation with Hamiltonian $\Hhat(\thetavec_i)$ at energy $E > 0$, normalized such that for every $i$,
\begin{align} \label{eq:unormalization}
    u^{(i)}_{\ell,E}(r) \underset{r\rightarrow\infty}{\longrightarrow}
     \frac{1}{p} \sin \bigg( pr - \ell \frac{\pi}{2} \bigg) + \frac{\Km_\ell^{(i)}(E)}{p} \cos \bigg( pr - \ell \frac{\pi}{2} \bigg)
     .
\end{align}
Here $p = \sqrt{2\mu E}$, the scattering wave function is decomposed as
\begin{align} \label{eq:partialwave}
    \langle \rvec | \psi_E(\thetavec_i)\rangle = \frac{u^{(i)}_{\ell,E}(r)}{r} Y_{\ell m}(\Omega_r)
    ,
\end{align}
and $\Km_\ell^{(i)}(E) = \tan \delta_\ell^{(i)}(E)$ is the partial-wave $K$ matrix element~\cite{taylor2006scattering} for $\Hhat(\thetavec_i)$ at energy $E$.

The KVP asserts that (also see the Supplementary Material (SM))~\cite{taylor2006scattering} 
\begin{align} \label{eq:KVPfunctional}
   \beta\bigl[\psitket\bigr] \equiv 
   \tau_{\text{trial}}
    - 2\mu
    \psitbra \Hhat(\thetavec) - E\psitket
    ,
\end{align}
subject to the radial part of $\langle\rvec\psitket$ being normalized as in Eq.~\eqref{eq:unormalization} but with $\Km_\ell^{(i)}(E)/p \rightarrow \tau_{\text{trial}}$, will be a stationary approximation to $[\Km_\ell(E)]_{\text{exact}}$
(i.e., it is accurate to second order in the difference of the exact and trial wave functions although not an upper bound in general).
The normalization condition for $\psitket$ is fulfilled if $\sum_{i=1}^{\Nb} c_i = 1$, which can be imposed with a Lagrange multiplier $\lambda$.
Substituting \eqref{eq:trialwfscatt} into \eqref{eq:KVPfunctional} with this constraint term and requiring the derivatives with respect to $c_i$ and $\lambda$ to be zero yields a simple matrix inversion problem with solution
\begin{align} \label{eq:solvingevc}
    c_i &= \sum_{j=1}^{\Nb} (\Delta\Utilde)^{-1}_{ij} \left( \frac{\Km_\ell^{(j)}(E)}{p}  - \lambda\right)
    , \\
    \label{eq:lambdadef}
   \lambda &= \frac{- 1 + \displaystyle\sum_{i,j=1}^{\Nb} (\Delta\Utilde)^{-1}_{ij}\frac{ \Km_\ell^{(j)}(E)}{p} }
         {\displaystyle\sum_{i,j=1}^{\Nb} (\Delta\Utilde)^{-1}_{ij}}
         ,
\end{align}
where
\begin{align} \label{eq:DeltaUtilde}
    \Delta\Utilde_{ij} \equiv {2\mu} 
      \langle\psi_E(\thetavec_i)|2\Vhat(\thetavec) - \Vhat(\thetavec_i) - \Vhat(\thetavec_j) | \psi_E(\thetavec_j)\rangle
      .
\end{align}
In obtaining Eq.~\eqref{eq:DeltaUtilde} we have used that $\bigl(\Hhat(\thetavec_i) - E\bigr)| \psi_E(\thetavec_i)\rangle = 0$ for every $i$.
Finally, the stationary approximation to the exact partial-wave $K$ matrix is
\begin{align} \label{eq:K_approx}
   [\Km_\ell(E)]_{\text{exact}} \approx
    \sum_{i=1}^{\Nb} c_i \Km_\ell^{(i)}(E)
    - \frac{p}{2}\sum_{i,j=1}^{\Nb} c_i \Delta\Utilde_{ij} c_j
    .
\end{align}
Thus the approximation is given by a weighted average of the $K$ matrices from the basis Hamiltonians with a correction term.

Note that the validity of the KVP relies only on the cancellation of $\delta\tau_{\text{trial}}$  with surface terms arising from the variation of 
$\psitbra \Hhat(\thetavec) - E\psitket$ when varying $ \beta\bigl[\psitket\bigr] $, which is satisfied by Coulomb, non-local, and complex potentials, as well as for coupled channels. (When the Coulomb potential is present, the asymptotic behavior of the scattering wave function is different from Eq.~\eqref{eq:unormalization}. For complex potentials, the $\langle\psi_E(\thetavec_i)|$ factors in Eqs.~\eqref{eq:KVPfunctional} and~\eqref{eq:DeltaUtilde}  need to be applied with time reversal~\cite{10.1143/PTPS.62.236, Barrett:1983zz}. See the discussion in the SM.) 
It is worth pointing out that any long-range potential in $\Hhat(\thetavec)$ independent of $\thetavec$, such as Coulomb, will cancel from $\Delta\Utilde_{ij}$ in Eq.~\eqref{eq:DeltaUtilde}
and one needs only to evaluate the matrix element within the range of the remaining potentials, which simplifies calculations. 
Also note that Eq.~\eqref{eq:DeltaUtilde} can be evaluated in momentum space or any other convenient basis.
More details on the derivation of Eqs.~\eqref{eq:solvingevc}--\eqref{eq:K_approx} are given in the SM.

As seen in Eqs.~\eqref{eq:solvingevc} to~\eqref{eq:DeltaUtilde}, the numerical effort is mainly composed of (a) constructing the $\Delta\Utilde$ matrix and (b) linear algebra operations with it. 
The computational cost in (a) can be significantly reduced by saving the $\thetavec$-independent pieces, which are also the most time-consuming ones, instead of computing them while sampling the parameter space. 
For (b), the small dimension space---$\Nb\sim 10$ in the following test examples---reduces both memory and time in the linear algebra calculations. In contrast, directly solving elastic scattering problems, such as those performed here using a R-matrix package~\cite{Descouvemont:2015xoa} (see the discussion below), involve operations with matrices having dimensions  of order $10^2$, which is much larger than $N_b$. Since the computational cost of optimized large-matrix manipulations including multiplication and inversion scale as the dimension to a power between 2 and 3, 
the cost reduction using EC can be significant with the per-sample cost in (a), when averaged over many sampling calculations, becoming negligible.
Nevertheless, the greatest advantage of EC will be for few-body scattering applications, for which the cost of direct calculations for large-scale sampling is prohibitive.

\begin{figure*}[tbh]
  \includegraphics{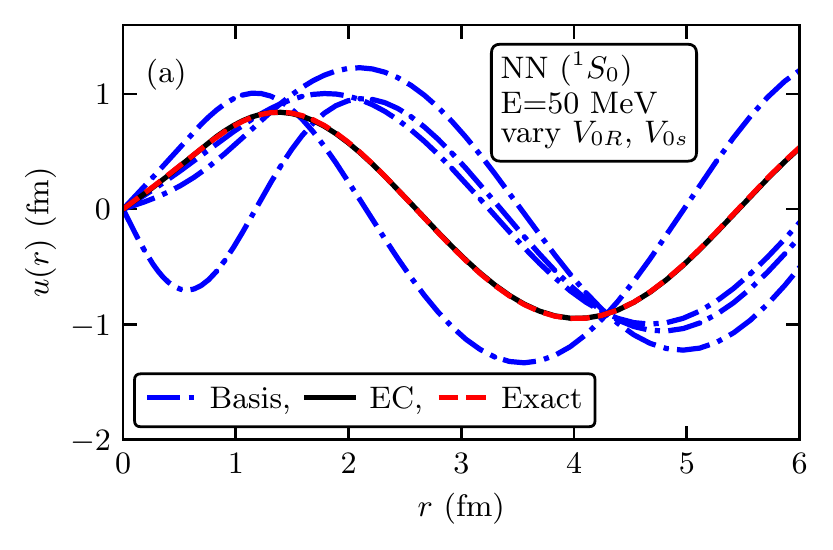}
  \includegraphics{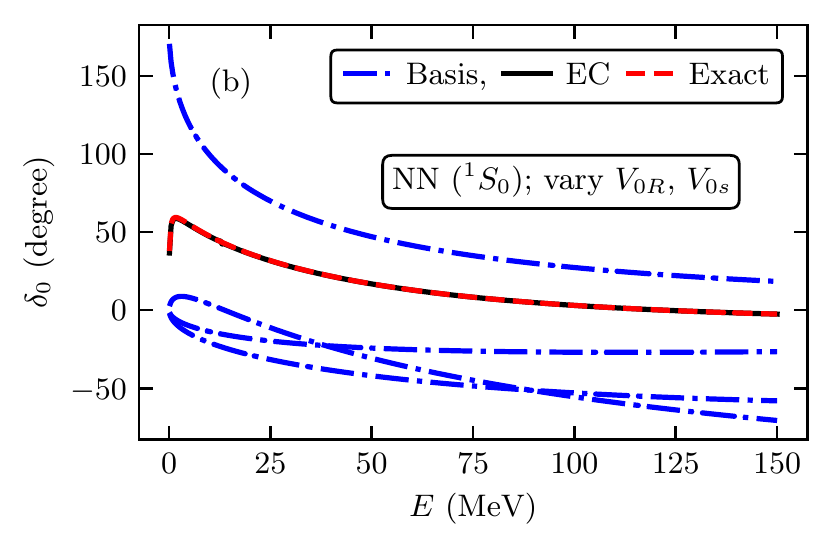}
    \caption{(a) Scattering wave functions for the Minnesota \singletS\ potential in Eq.~\eqref{eq:Minnesota_1S0} with $E=50$ MeV (in the CM frame). 
    The dot-dashed curves are for four choices of $\thetavec_i = \{V_{0R},V_{0s}\}$ that comprise the \EVC\  trial basis, the dashed curve is for the exact values from Ref.~\cite{THOMPSON197753}, and the solid curve is the \EVC\  prediction. 
    The curves have a common crossing point at the value of $r$ where the second term in Eq.~\eqref{eq:unormalization} is zero.
    (b) Scattering phase shifts for the same parameter sets and the \EVC\  prediction.} 
  \label{fig:NN2dim1S0_NB_4_E_50_WF}
\end{figure*}

The matrix $\Delta\Utilde$ to be inverted may be expected to be increasingly ill-conditioned as the basis size $\Nb$ increases.
Even for conventional applications of the KVP, there will be ill-conditioning issues for certain values of $E$, giving rise to so-called ``Kohn anomalous singularities''~\cite{PhysRev.124.1468, nesbet1980variational}.
The often-recommended remedy is to use a complex formulation (involving the $S$ matrix rather than the $K$ matrix), which mostly avoids the problem~\cite{doi:10.1063/1.454462,PhysRevA.40.6879,ADHIKARI1992415,RevModPhys.68.1015}.
Here we also have ill-conditioning, but at all $E$ for sufficiently large ${\Nb}$.
We find, however, that a simple regularization of the smallest singular values of $\Delta\Utilde$ is sufficient to ameliorate the ill-conditioning~\cite{Neumaier98solvingill-conditioned,engl1996regularization}.
This can be done by adding a small value to the diagonal of $\Delta\Utilde$ (called a nugget in this context, but cf.\ Tikhonov regularization~\cite{Neumaier98solvingill-conditioned,engl1996regularization}) or by using the pseudo-inverse in Eq.~\eqref{eq:lambdadef}.
Because we can accurately calculate test results, we can verify the efficacy of the regularization. In the following calculations, the nugget is chosen to be between $10^{-10}$ and $10^{-8}$  to optimize---by hand---those \EVC\ estimations with an ill-conditioning problem. 
Kohn anomalous singularities are still present at isolated energies, but are only noticeable on a fine $E$ mesh. For applications of emulators to sampling this should not be an issue; if necessary they can be mitigated by comparing different results from changing the basis size by one, as the position of these unphysical singularities will move.

\begin{figure}[tbh]
  \includegraphics{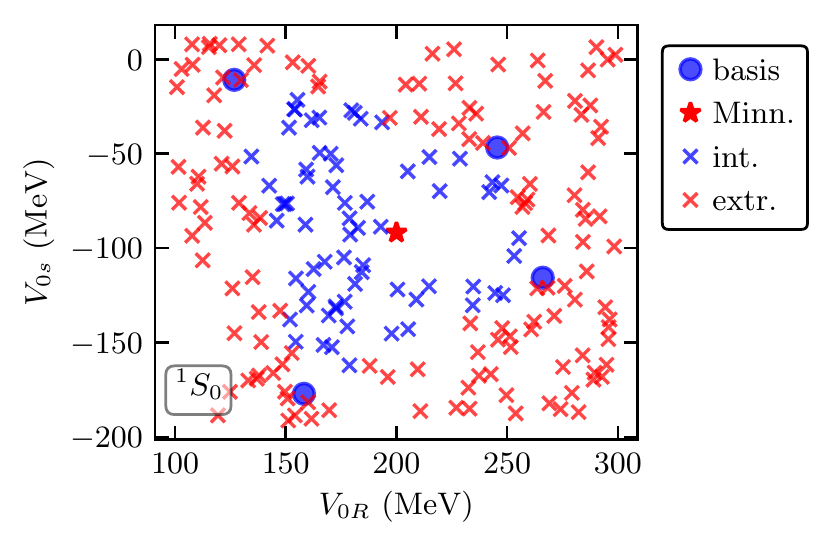}
    \caption{Sampled points in the parameter space for the Minnesota potential in the \singletS\ channel.
    The best parameter set from Ref.~\cite{THOMPSON197753} is a star, four values for the \EVC\  trial basis are circles, and the crosses are test point, which are either interpolations (blue) or extrapolations (red). }
  \label{fig:NN2dim1S0_pts_NB_4}
\end{figure}


\section{\EVC\  for a model of NN scattering} \label{sec:NN_scattering}

We use the "Minnesota potential"~\cite{THOMPSON197753}, which was developed to reproduce $^{1}S_{0}$ and $^{3}S_{1}$ nucleon-nucleon (NN) scattering phase shifts with a simple functional form, as a test example to explore the application of \EVC\ for scattering. 
The potential is a sum of local Gaussian terms, without Coulomb interaction or coupled channels. 
Each S-wave channel has a repulsive short-range term and an attractive term with longer range:
\begin{align} \label{eq:Minnesota_1S0}
  V_{^{1}S_{0}} (r) &\equiv  V_{0R} e^{-\kappa_R r^2} + V_{0s}  e^{-\kappa_s r^2}     \ ,  
  \\    \label{eq:Minnesota_3S1}
  V_{^{3}S_{1}} (r) & \equiv V_{0R}  e^{-\kappa_R r^2} + V_{0t}  e^{-\kappa_t r^2}    
\end{align} 
The best values from Ref.~\cite{THOMPSON197753} are $200.$, $-91.85$, and $-178$ MeV for $V_{0R}$, $V_{0s}$, and $V_{0t}$, and $1.487$, $0.465$, and $0.639$ $\mathrm{fm}^{-2}$ for $\kappa_R$, $\kappa_s$ and $\kappa_t$.

To illustrate how \EVC\  works, values are chosen for $\thetavec_i = \{V_{0R},V_{0s}\}$ for $i=1$ to 4, to form a trial basis for \EVC\  calculations in the \singletS\ channel.
These points in the parameter space are $(0., -291.85)$, $(100.,  8.15)$, $(300. , -191.85)$, and $(300. , 8.15)$. 
Figure~\ref{fig:NN2dim1S0_NB_4_E_50_WF}(a) shows the scattering wave functions at $E=50$ MeV (in the center-of-mass (CM) frame) from the four basis potentials (blue dot-dashed lines), the exact wave function corresponding to the best value parameters at the same energy (red dashed), and the wave function from \EVC\  based on the four-potential basis (black solid line). 
It is evident that with four basis elements 
the \EVC\  wave function agrees very well with the exact wave function. 
Figure~\ref{fig:NN2dim1S0_NB_4_E_50_WF}(b) shows the corresponding phase shifts.
Again, the exact result is very well reproduced by the \EVC\  prediction, even though the wave functions and phase shifts of individual basis elements are significantly different.

\begin{figure*}[t]
  \includegraphics[width=\textwidth]{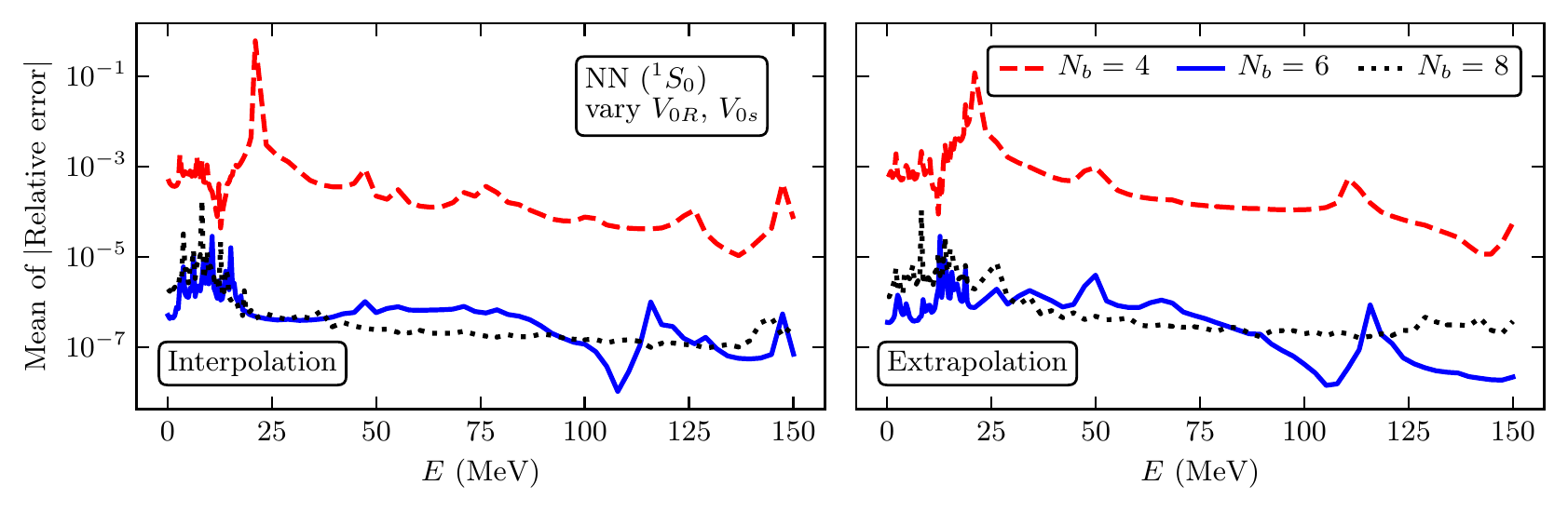}
  
    \caption{Relative errors between \EVC\  predictions with $\thetavec_i = \{V_{0R}, V_{0s}\}$ and direct calculations of $p/\Km_\ell(E) = p\cot\delta(E)$ for the Minnesota potential in the \singletS\  channel.
    This is the mean of the errors for (a) interpolated and (b) extrapolated parameter sets as shown in Fig.~\ref{fig:NN2dim1S0_pts_NB_4}. The size of the nugget used for inverting $\Delta\Utilde$ is $10^{-10}$ here.}
  \label{fig:NN2dim1S0}
\end{figure*}

\begin{figure*}[thb]
  \includegraphics[width=\textwidth]{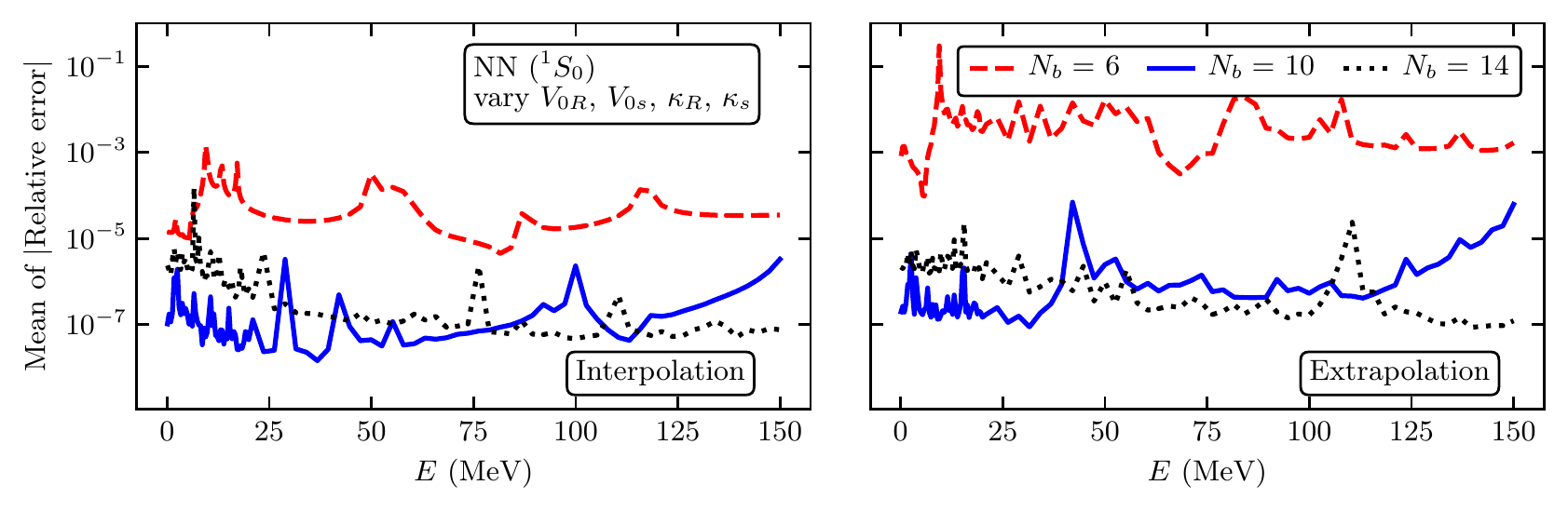}
    \caption{Same as Fig.~\ref{fig:NN2dim1S0} but for a four-dimensional parameter space with $\thetavec_i = \{V_{0R},\kappa_R,V_{0s},\kappa_s\}$. The nugget is set to $10^{-9}$ here.}
  \label{fig:NN4dim1S0}
\end{figure*}

Next we make a more global study with the same potential.
For each channel, we vary the two potential strengths by $\pm 100$ MeV about the best values, and scan the 2-dimensional parameter space by comparing the \EVC\ -phase shift with the exact phase shift. 
The values for the $\thetavec_i = \{V_{0R},V_{0s}\}$ parameters are randomly drawn using Latin-hypercube sampling~\cite{doi:10.1080/01621459.1993.10476423}, as used in \EVC\  bound-state studies~\cite{Konig:2019adq,Ekstrom:2019lss}. 
A range of basis sizes ${\Nb}$ have been explored. 
For those parameter values, we compute scattering phase shifts and wave functions by directly solving the  Schr\"odinger equation using an $R$-matrix package~\cite{Descouvemont:2015xoa}, which serves as input for the subsequent \EVC\  calculations using Eqs.~\eqref{eq:solvingevc}--\eqref{eq:DeltaUtilde}. 
To explore the predictive power of the \EVC, we randomly sampled 200 points from the two-dimensional space, and for each made \EVC\  predictions as well as direct calculations using the $R$-matrix package, whose phase-shift calculation, as we checked, has precision---i.e., relative error---better than $10^{-8}$ with the order of $10^2$ mesh points used therein.  
Comparing these results indicates the  accuracy of the \EVC\  emulator. 

An example of the parameter sets for this comparison protocol is shown in Fig.~\ref{fig:NN2dim1S0_pts_NB_4}, where the sampled points in the $V_{0R}$--$V_{0s}$ parameter space (for the \singletS\ channel) are shown.
The trial basis points ($\Nb=4$) are blue circles, the tested sample points are blue crosses if within the convex hull of the basis points (for these the \EVC\  calculations are considered to be interpolations) and otherwise are red crosses (these \EVC\  calculations are extrapolations), and finally the best-value point is a red star. 

In Fig.~\ref{fig:NN2dim1S0}, the mean values of the relative error (in absolute value) of the \EVC\  calculations for the interpolated sample points (left panel) and the extrapolated points (right panel) are plotted against the scattering energy $E$ (in the CM frame) for three calculations using $\Nb=4$, 6, and 8 basis elements.
(The errors are in the value of $p/\Km_\ell(E) = p\cot\delta(E)$. Since the relative error is tiny here, other functions of $\delta$ have almost the same relative errors.)
With a basis size of 4, the \EVC\  calculation can reproduce the phase shift to better than 0.1 percent at almost all energies.
The accuracy improves to be better than $10^{-4}$, and for most energies it reaches $10^{-6}$, with $\Nb=6$. 
For $\Nb=8$, the $\Delta\Utilde$ matrix becomes ill-conditioned, but after regularizing the small singular values by adding a nugget ($10^{-10}$) to the diagonal of this matrix when computing the matrix inversion in Eqs.~\eqref{eq:solvingevc} and~\eqref{eq:lambdadef}, the accuracy of these calculations is comparable to the $\Nb=6$ case. 
We also computed the standard deviations of the absolute value of the relative errors, and found them to be similar in size to the mean values. 
It is interesting to note in this case that \EVC\  works equally well for interpolated and extrapolated points. 

The major spikes in these plots and the following figures show that in a subset of EC calculations, the combination of potential parameter values and the energy can get close enough to a Kohn anomalous singularity that the corresponding relative error is dramatically larger than the errors of the nearby points. 
However for a  typical application of emulators, we expect low probability for such fine tuning. 
For  reference, in Fig.~\ref{fig:NN2dim1S0}, there are 200 uniformly sampled in a two-dimensional parameter space with a 1\,MeV mesh in $E$. Most importantly, the locations of spikes and thus the singularities vary among the different EC basis sets,
so detecting and mitigating them is straightforward.

To explore a higher-dimensional parameter set, we vary both the potential strength ($\pm 100$ MeV about the best values) and the two Gaussian widths $\kappa_R$ and $\kappa_s$ within a $\pm 50\%$ range about their best values.  
So now $\thetavec_i = \{V_{0R},\kappa_R,V_{0s},\kappa_s\}$.%
\footnote{Note that because the $\kappa$ parameters do not appear linearly in the Hamiltonian, one can no longer make a single set of matrix elements calculations for all of the test parameter sets. In other contexts this might be a relevant computational disadvantage.}
For this demonstration, we uniformly sample 1000 test points within the four-dimensional parameter space. 
Figure~\ref{fig:NN4dim1S0} shows the parallel error information to Fig.~\ref{fig:NN2dim1S0}, with a nugget of $10^{-9}$ size.  
For the interpolated points, the accuracy improves from $10^{-3}$--$10^{-4}$ to $10^{-6}$ or better as $\Nb$ increases from 6 to 10. 
For $\Nb=14$, the ill-conditioning issues require the use of a nugget but its accuracy is similar to $\Nb=10$. 
The results for the extrapolated parameter sets are worse than the interpolated results for $\Nb = 6$, but become as accurate with $\Nb=10$ and larger. 
Again, the standard deviations of the relative errors are comparable to their mean values. 
The parallel results for the \tripletS\ channel are similar for a large enough trial basis (see the SM). 

\section{Other examples: $p$--$\alpha$ and $\alpha$--P\lowercase{b}} 
\label{sec:other_examples}

\begin{figure}[tbh]
  \includegraphics[width=0.48\textwidth]{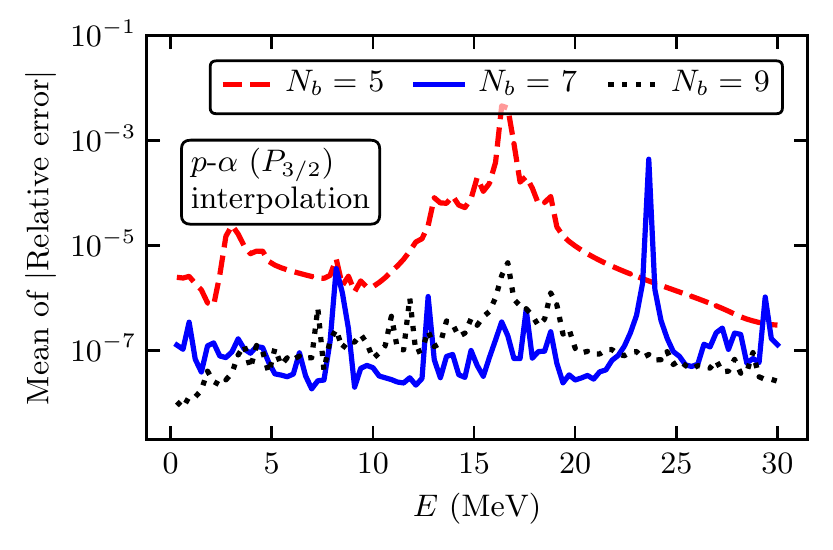}
    \caption{The relative errors for $\tan\delta(E)$ in the $p$--$\alpha$ $P_{3/2}$ channel in the two-dimensional space $\thetavec_i = \{V^{(0)}_{p\alpha,1},\beta_1\}$. The nugget is set to $10^{-8}$ here. }
  \label{fig:palpha2dimP3half}
\end{figure}

To explore the effectiveness of \EVC\  for non-local potentials, the inclusion of a Coulomb potential, and for higher-partial waves, we use proton--$\alpha$ scattering in the $S_{1/2}$ and $P_{3/2}$ channels as examples, with the non-local potential~\cite{Ali:1984ds}: 
\begin{align}
  V_\angL(r', r) = V^{(0)}_{p\alpha,\angL} \; r'^\angL\,  r^\angL  e^{-\beta_\angL \; (r' +r ) } . 
\end{align}
The best values are: $V^{(0)}_{p\alpha,0} = -168.28$ MeV, $\beta_0= 0.8~\mathrm{fm}^{-1}$, $V^{(0)}_{p\alpha,1} = -291.26$ MeV, and $\beta_1= 1.25~\mathrm{fm}^{-1}$. 
The Coulomb potential takes the point-charge form.
For each of the two channels,  we vary both the potential strengths $V^{(0)}_{p\alpha,\angL}$ around its best values $\pm 100$ MeV and the width parameters $\beta_\angL$ around its best values $\pm 50\%$. 
As a representative case,
the relative errors for interpolated points in the $P_{3/2}$ channel are plotted in Fig.~\ref{fig:palpha2dimP3half} for several basis sizes (additional plots for the $S_{1/2}$ channel are given in the SM). The nugget for both channels is set to be $10^{-8}$. 
The performance of \EVC\  is again excellent except at some isolated energies, and these exceptions are not at the same energies for different basis sizes.

The Kohn variational approach also applies to complex potentials, which are extensively used in optical potentials for nuclear scattering and reactions. 
To test the \EVC\  for these applications, we use a  Wood-Saxon optical potential constructed for describing $\alpha$--$^{208}\mathrm{Pb}$ low-energy scattering~\cite{GOLDRING1970465}: 
\begin{align}
    V(r) =  V_{0}\; f(r, R_R, a_R) + i W_{0} \; f(r, R_I, a_I) , 
\end{align}
with $f(r, R, a) \equiv \left(1 + \exp{(r-R)/a}\right)^{-1}$. We take 
$V_{0}= -100$ MeV, $W_{0}=-10$ MeV, $R_R =R_I = 8.36$ fm, and $a_R=a_I= 0.58$ fm as the best value~\cite{Descouvemont:2015xoa}. The Coulomb potential is simplified as for point charges~\cite{Descouvemont:2015xoa}.

\begin{figure}[thb]
  \includegraphics[width=0.48 \textwidth]{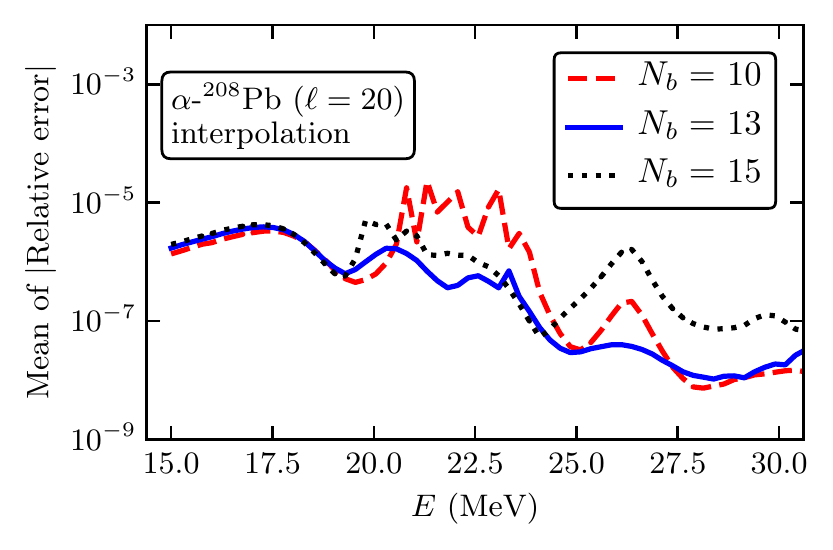}
  
    \caption{The relative errors for $\tan\delta(E)$ in the $\alpha$-$^{208}\mathrm{Pb}$ $\ell=20$ channel in the two-dimensional space $\thetavec_i = \{V_0,W_0\}$. Note that $\delta$ is complex. The nugget is set to $10^{-10}$ here.}
  \label{fig:alphaPb2diml20}
\end{figure}

We  vary the $V_0$ and $W_0$ parameters  in different partial waves (so $\thetavec_i = \{V_0,W_0\}$) by $\pm 50\%$ around their best values~\cite{Descouvemont:2015xoa}.  
In Fig.~\ref{fig:alphaPb2diml20}, the size of relative errors for the $\ell=20$ channel is shown as a representative example. The results for $\ell=0$ are shown in the SM.  (Note that the scattering phase shift is complex here. The vertical axis of the plot is for the modulus of the relative error.) 
The lower end of the energy range is chosen such that the Sommerfeld parameter $\eta$ is less about $10$, because the numerical calculation of Coulomb functions in the $R$-matrix package becomes unreliable for larger values~\cite{BARNETT1982147, Descouvemont:2015xoa}. 
The upper end of the energy range is chosen to match Ref.~\cite{GOLDRING1970465}. 
The nugget used in $\Delta\Utilde$'s inversion is set to $10^{-10}$ in both $\angL=20$ and $\angL=0$ calculations.
With 10 basis elements, the relative accuracy for interpolated points is no worse than $10^{-4}$, while increasing $\Nb$ further improves it to $10^{-5}$ or better. 
Again, the errors for interpolated and extrapolated points are similar, and the standard deviations are similar in size to the mean values.

Based on these results for $p$-$\alpha$ and $\alpha$-$\mathrm{Pb}$ scattering, we expect  \EVC\ could play an important role in fitting potential parameters for nuclear scattering and coupled-channel reactions. 

\section{Summary and outlook} 

We have extended the eigenvector continuation method to scattering using the Kohn variational  principle.
The \EVC\ enables accurate calculations of observables for any parameter set $\thetavec$ given calculations of scattering wave functions and $K$-matrix elements from a limited number $\Nb$ of parameter sets $\thetavec_i$.
Unlike the bound-state application of \EVC, for scattering the KVP does not give an upper bound to observables but is only guaranteed to give stationary results.
Nevertheless, for good trial functions the KVP has been demonstrated in the literature to give accurate results for a wide range of applications~\cite{nesbet1980variational}.
An \EVC\ basis provides a very effective trial function and its application to the KVP is simple, involving only the inversion of the matrix defined in Eq.~\eqref{eq:DeltaUtilde}.
Issues of ill-conditioning with increasing basis size are successfully treated with simple regularizations.

Here we have provided representative results from a wide range of tests of the \EVC\ for scattering using model problems.
These include multi-dimensional parameter sets, both local and non-local potentials, charged-particle scattering, and complex optical potentials. 
In all cases shown here and in all our other tests to date, the \EVC\ is found to be effective with moderate basis sizes both for interpolated and extrapolated parameter sets.
We are working to formulate a robust uncertainty quantification and to develop a procedure for determining the optimal regularization parameter for ill-conditioning, which has thus far been fixed empirically.

The success of the \EVC\ enables the development of efficient emulators for scattering.
In subsequent work we will demonstrate the application to coupled channels in both coordinate-space and momentum-space (which is a straightforward generalization of the presentation here) and set up the application to Nd scattering~\cite{,Kievsky:1997bg,Kohn:1948zz}. 
It would be also interesting to apply our method to fit an NN potential to the NN energy spectra from Lattice QCD calculations, since the eigenenergies and phase shifts are directly connected.

\acknowledgments{
We are grateful for stimulating discussions with C. Greene and with members of the BAND Framework project~\cite{BAND_framework}.
Useful feedback on the manuscript was provided by S.~K\"onig, D. Lee, and J.~Melendez.
This work was supported in part by the National Science Foundation
under Grant No.~PHY--1913069 and the CSSI program under award number OAC-2004601 (BAND Collaboration), and by the NUCLEI SciDAC Collaboration under
Department of Energy MSU subcontract RC107839-OSU.}


%

\clearpage
\onecolumngrid
\begin{center}
  \textbf{\large Supplementary Material for  Eigenvector continuation for scattering }\\[.4cm]
  R.~J. Furnstahl,$^1$ A.~Garcia,$^1$ P.~J. Millican,$^1$ Xilin Zhang,$^1$ \\[.1cm]
  $^1${\itshape Department of Physics, The Ohio State University, Columbus, Ohio 43210, USA} 
 \end{center}

\twocolumngrid

\setcounter{equation}{0}
\setcounter{figure}{0}
\setcounter{table}{0}
\setcounter{section}{0}
\setcounter{page}{1}
\makeatletter
\renewcommand{\theequation}{S\arabic{equation}}
\renewcommand{\thefigure}{S\arabic{figure}}

\maketitle

\section{Kohn variational method }

To have a self-contained presentation, we briefly review the KVP~\cite{Kohn:1948zz,taylor2006scattering} for single-channel scattering. 
The generalization to coupled-channel processes is straightforward~\cite{Kohn:1948zz,10.1143/PTPS.62.236}.     
For a specified partial wave $\ell$ with local potential $V(r)$,  the KVP stationary functional \eqref{eq:KVPfunctional}  with the trial radial wave function $\utrial$ (we suppress the dependence on $\ell$ and energy $E$) reduces to~\cite{taylor2006scattering}
\begin{align} \label{eq:beta_def}
    \beta[\utrial] = \tau_{\text{trial}} - \int_0^\infty dr\, \utrial(r) D \utrial(r)
    .
\end{align}
Here, $p$ is the asymptotic momentum ($p = \sqrt{2\mu E}$) and 
\begin{align} \label{eq:D_def}
    D \equiv - \frac{d^2}{dr^2} + \frac{\ell(\ell+1)}{r^2} + U(r) - p^2
    ,
\end{align}
with $U(r) \equiv 2\mu V(r)$. 
The value of $\tau_{\text{trial}}$ is extracted from the asymptotic behavior of the trial wave function, 
\begin{align} \label{eq:asymptotic}
    \utrial(r) \underset{r\rightarrow\infty}{\longrightarrow}
      \frac{1}{p}\sin(pr - \frac12 \ell \pi) +
      \tau_{\text{trial}} \cos(pr - \frac12 \ell \pi)
      ,
\end{align}
where the sine term sets the normalization.
This normalization convention goes hand-in-hand with the particular form of the $\beta[\utrial]$ functional, which needs to be modified for other normalizations. 

For the exact radial function $u_{\text{exact}}$, we have
\begin{align} \label{eq:exact_zero}
    D u_{\text{exact}}(r) = 0
    ,
\end{align}
and therefore from Eqs.~\eqref{eq:asymptotic} and~\eqref{eq:unormalization} with $\tau_{\text{trial}}\rightarrow\tauex$, 
\begin{align} \label{eq:tau_exact}
   \beta[\uexact] = \frac{1}{p}[\Km_\ell(E)]_{\text{exact}}
   = \frac{1}{p} [\tan\delta_\ell(E)]_{\text{exact}}
   .
\end{align}
To see that $\beta$ is a stationary functional, we write 
\begin{align}
    \utrial(r) = \uexact(r) + \delta u(r)
\end{align}
and substitute into $\delta\beta = \beta[\utrial] - \beta[\uexact]$. 
Using Eq.~\eqref{eq:exact_zero} and keeping only to first order in $\delta u$, we obtain
\begin{align}
   \delta\beta = \delta\tau - \int_0^\infty dr\,\uexact(r) D \delta u(r)   + \mathcal{O}(\delta u^2) .
\end{align}
We want to act $D$ to the left to take advantage of Eq.~\eqref{eq:exact_zero}, which requires partially integrating the deriatives in $D$.
This yields only surface terms, to which we can use the asymptotic forms:
\begin{align}
    \delta\beta = \delta\tau +  
     (\uexact\delta u' - \uexact'\delta u) \Bigr|_0^\infty 
     + \mathcal{O}(\delta u^2) .
\end{align}
The lower limit does not contribute because
\begin{align}
    \uexact(0) = \utrial(0) = \delta u(0) = 0 \;,
\end{align}
while the upper limit, after using Eq.~\eqref{eq:asymptotic} for $\uexact$ and $\utrial$, yields $-\delta\tau[\sin^2(pr - \frac12\ell\pi) + \cos^2(pr - \frac12\ell\pi)] = -\delta \tau$ (with two other terms canceling), so in the end
\begin{align}
    \delta\beta = 0 + \mathcal{O}(\delta u^2)
    .
\end{align}
Thus, the functional $\beta[\utrial]$ approximates $\tauex$ at its stationary point up to $\mathcal{O}\bigl(\delta u^2\bigr)$.

If the long-range Coulomb potential is present, the asymptotic behavior of the radial basis functions and of $\utrial$ differs by the argument in the sine and cosine functions, namely,  
\begin{align}
\sin(pr - \frac12 \ell \pi)  \rightarrow \sin(pr - \frac12 \ell \pi - \eta \ln{2 p r} + \sigma_\angL)  \ , 
\end{align}
and similarly for the $\cos$ function. Here $\eta$ is the Sommerfeld parameter and $\sigma_\angL$ the pure Coulomb phase shift~\cite{taylor2006scattering}. Then the phase shift $\delta_\ell$ would be the shorter-range interaction-induced phase shift, i.e., the total phase shift with $\sigma_\ell$ subtracted. (Note that the total phase shift here is measured with respect to the incoming and outgoing spherical waves in the form of $\exp\left[\pm i (p r -\eta \ln{2pr})\right]$.)

For a general non-local potential, the above derivation still holds, except that the integral involving $V(r)$ in Eq.~\eqref{eq:beta_def} needs to be changed to a double integration, i.e., $\int dr dr' \utrial(r) V(r, r') \utrial(r')$. 
It should be emphasized that the potential in the function may be complex in general, but in this case no complex conjugation is to be applied to $\utrial$ in the integral~\cite{10.1143/PTPS.62.236}. 
The applicability of KVP for complex potentials is important for this approach to be used in optical potential model fitting. 

The generalization to coupled channels can be found e.g., in Ref.~\cite{Barrett:1983zz, 10.1143/PTPS.62.236}. 
The central step involves finding that the variation of the integral in the definition of the $\beta$ functional comes from the end points/surface terms of the integral (using Green's theorem), which exactly cancel the variation of the first term in Eq.~\eqref{eq:beta_def}, which now turns into the K-matrix $\Km$~\cite{Barrett:1983zz}.   

\section{Adapting \EVC\  to scattering}
 \label{sec:adapting_EVC}

To adapt \EVC\  to the scattering problem, we do not solve an energy eigenvalue problem but instead we find solutions at a specified scattering energy $E =  p^2 / 2\mu$.
We still have a set of $\Nb$ Hamiltonians specified by $\{\thetavec_i\}$ but now for each one we find the scattering solution for energy $E$ and the corresponding phaseshift.
This will be our basis.
This identifies the $\tau_i = \Km_\ell^{(i)}/p$ value in Eq.~\eqref{eq:unormalization} for each basis wave function.
A trial wave function as in Eq.~\eqref{eq:trialwfscatt}, which we write as
\begin{align} \label{eq:utrial_EVC}
    \utrial(r) = \sum_i c_i u_i(r; E)  ,
\end{align}
with each $u_i$ normalized according to Eq.~\eqref{eq:unormalization}, will have the asymptotic form
\begin{align}
    \utrial(r) \underset{r\rightarrow\infty}{\longrightarrow}
     & \left(\sum_{i=1}^{N} c_i\right) \frac{1}{p}\sin(pr - \frac12 \ell \pi)  \notag \\ 
     & + \left(\sum_{i=1}^{N} c_i \tau_i \right) \cos(pr - \frac12 \ell \pi)
     ,
\end{align}
because the $u_i$ all have the same $p$.
Matching, we find the constraint:
\begin{align} \label{eq:constraint}
    \sum_{i=1}^{N} c_i = 1 .
\end{align}

So we substitute \eqref{eq:utrial_EVC} into \eqref{eq:beta_def} and require $\beta$ to be stationary with respect to the $c_i$s, subject to the constraint \eqref{eq:constraint}, which we incorporate using a Lagrange multiplier $\lambda$. The functional is 

\begin{widetext}
\begin{align} \label{eq:beta_EVC}
    \beta[\utrial] &= \sum_j c_j \tau_j(E) - 
        \sum_{jk} c_j c_k \int_0^\infty dr\,
        u_j(r; E)  
        \left[ -\frac{d^2}{dr^2} + \frac{l(l+1)}{r^2}
           + 2\mu V(r;\thetavec) - p^2\right] u_k(r;E) \notag \\
      &= \sum_j c_j \tau_j - \sum_{jk} c_j c_k \int_0^\infty dr\,
      u_j(r;E) (2\mu)  \Bigl[V(r;\thetavec) - V_k(r)\Bigr]
        u_k(r;E) .
\end{align}
\end{widetext}
To get the second line we have added and subtracted $V_k(r)$ inside the
integral and used that
each of the $u_j(r)$ wave functions are eigenstates with their corresponding
$V_j$ but all the same $E$.

We define
\begin{align}
    \Delta U_{jk} \equiv \int_0^\infty dr\,
      u_j(r;E) (2\mu)  \Bigl[V(r;\thetavec) - V_k(r)\Bigr]
        u_k(r;E) . \label{eq:deltaUdef}
\end{align}
Note that this is not a symmetric matrix. Then the functional to make stationary is:
\begin{align} \label{eq:tauex}
    \tauex \approx
    \beta[\utrial] = \sum_j c_j\tau_j - \sum_{jk} c_j \Delta U_{jk} c_k 
    .
\end{align}
We want to be this stationary under the constraint that the sum of the $c_j$ coefficients is one. So
\begin{align}
    \frac{\partial}{\partial c_i} &
    \biggl[ \sum_j c_j\tau_j - \sum_{jk} c_j c_k \Delta U_{jk} 
    - \lambda \Bigl(\sum_j c_j - 1 \Bigr)
    \biggr] = 0 , \\
        \frac{\partial}{\partial \lambda} &
    \biggl[ \sum_j c_j\tau_j - \sum_{jk} c_j c_k \Delta U_{jk} 
    - \lambda \Bigl(\sum_j c_j - 1 \Bigr)
    \biggr] = 0  .
\end{align}
The second equation just gives us back the constraint.  The first line, for each $i$, gives
\begin{align}
    \tau_i - c_k \Delta U_{ik} - c_j \Delta U_{ji} - \lambda = 0
\end{align}
or
\begin{align}
  \sum_j \bigl( \Delta U^{\intercal} + \Delta U \bigr)_{ij} c_j
   &\equiv \sum_j \Delta\Utilde_{ij} c_j \\
   &= \tau_i - \lambda
   ,
\end{align}
which defines $\Delta\Utilde_{ij}$ as in Eq.~\eqref{eq:DeltaUtilde}.
Finally,
\begin{align} \label{eq:c_coeff_eq}
    c_j = \sum_i (\Delta\widetilde  U)^{-1}_{ji}
        (\tau_i - \lambda)
        ,
\end{align}
in agreement with Eq.~\eqref{eq:solvingevc}.

We can identify $\lambda$ by summing this last equation over $j$ and setting it equal to 1, then solving for $\lambda$:
\begin{align} \label{eq:lambda_eq}
    \lambda = \frac{\sum_{ij} (\Delta\widetilde  U)^{-1}_{ji}\tau_i - 1}{\sum_{ij} (\Delta\widetilde  U)^{-1}_{ji}}  ,
\end{align}
in agreement with Eq.~\eqref{eq:lambdadef}.
To get our estimate of $\tauex = [\Km_\ell(E)]_{\text{exact}}/p$ as in \eqref{eq:K_approx}, we substitute from \eqref{eq:c_coeff_eq} and \eqref{eq:lambda_eq} into \eqref{eq:tauex}.
Note that the first term on the right side of Eq.~\eqref{eq:K_approx} is zeroth order in $\delta u$ while the correction term is first order.

\section{Additional results}

\begin{figure*}[tbh]
  \includegraphics[width=0.85\textwidth]{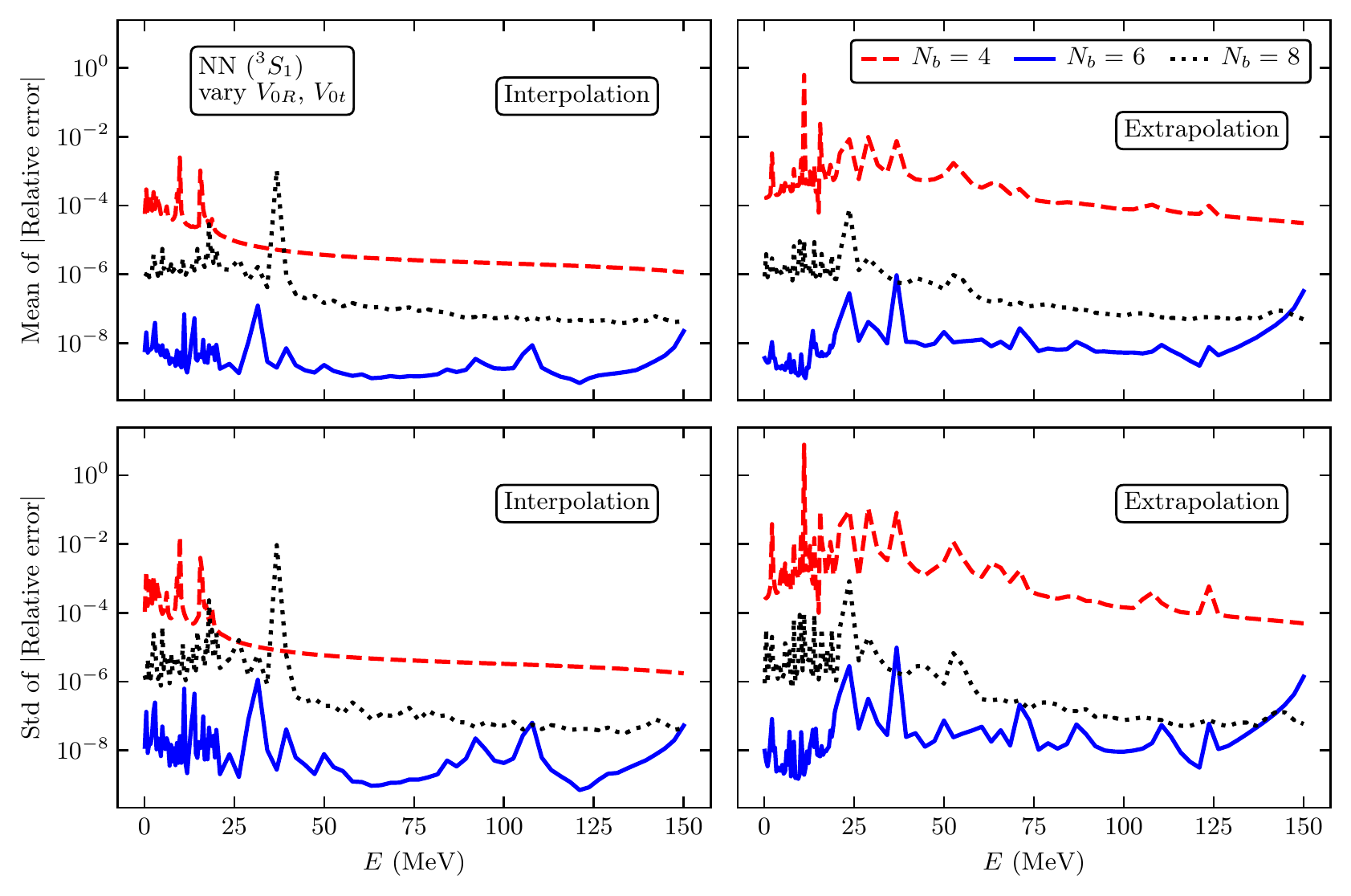}
  
    \caption{Relative errors between \EVC\  predictions and direct calculations of $p/\Km_\ell(E) = p\cot\theta(E)$ for the Minnesota potential in the \tripletS\  channel. To regularize the $\Delta\Utilde$ matrix for its inversion, the nugget is set to $10^{-9}$ here. 
    The top plots show the means of errors for sampled parameter sets in Fig.~\ref{fig:NN2dim1S0_pts_NB_4} while the bottom plots show the standard deviations of these errors.
    }
  \label{fig:NN2dim3S1}
\end{figure*}

\begin{figure*}[tbh]
  \includegraphics[width=0.85\textwidth]{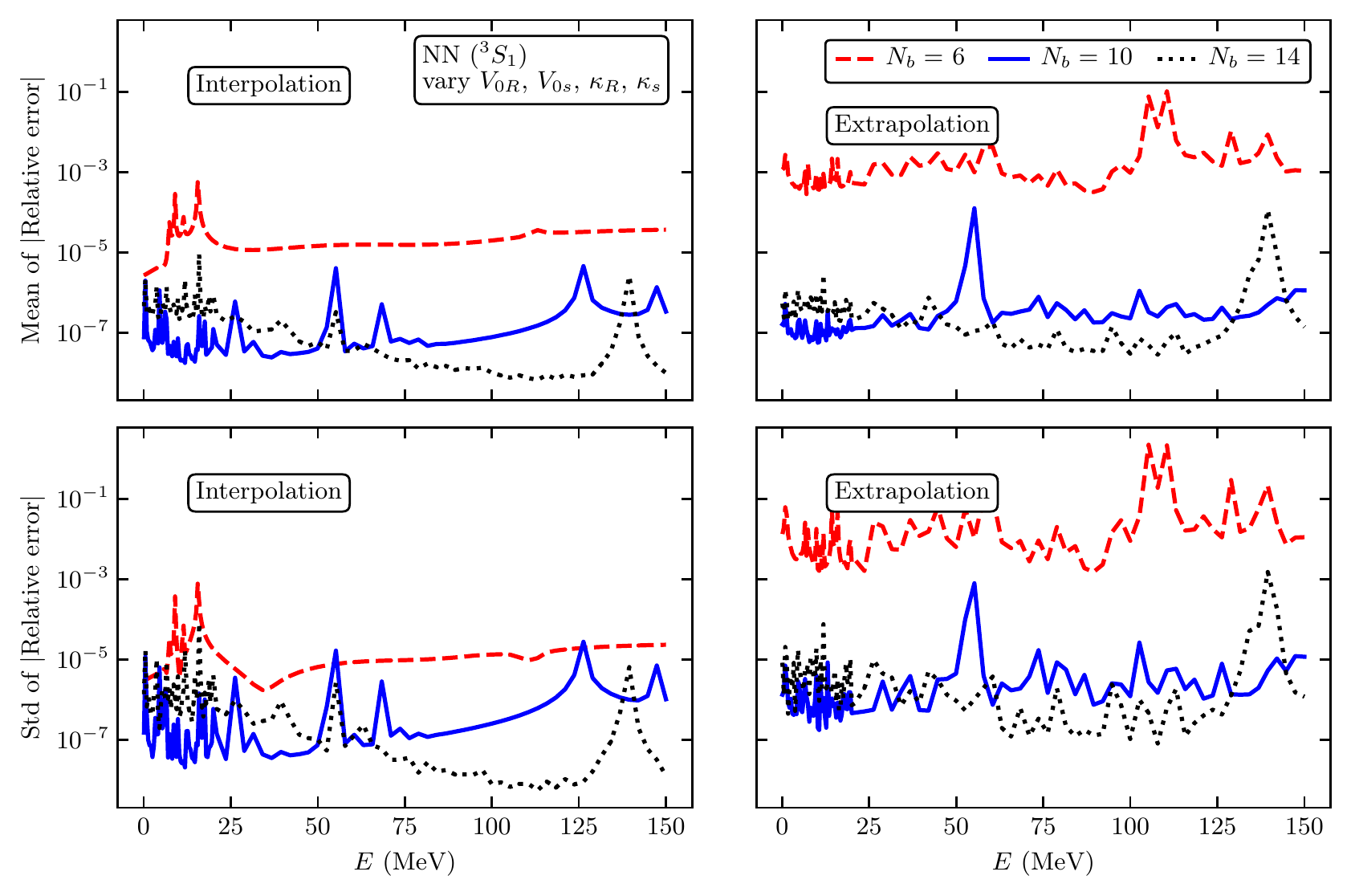}
  
    \caption{Same as Fig.~\ref{fig:NN2dim3S1} but for the 4-dimensional parameter space $\thetavec_i = \{V_{0R},\kappa_R,V_{0s},\kappa_s\}$. The nugget for $\Delta\Utilde$'s inversion is set to $10^{-8}$. }
  \label{fig:NN4dim3S1}
\end{figure*}

\begin{figure*}[tbh]
  \includegraphics[width=0.85\textwidth]{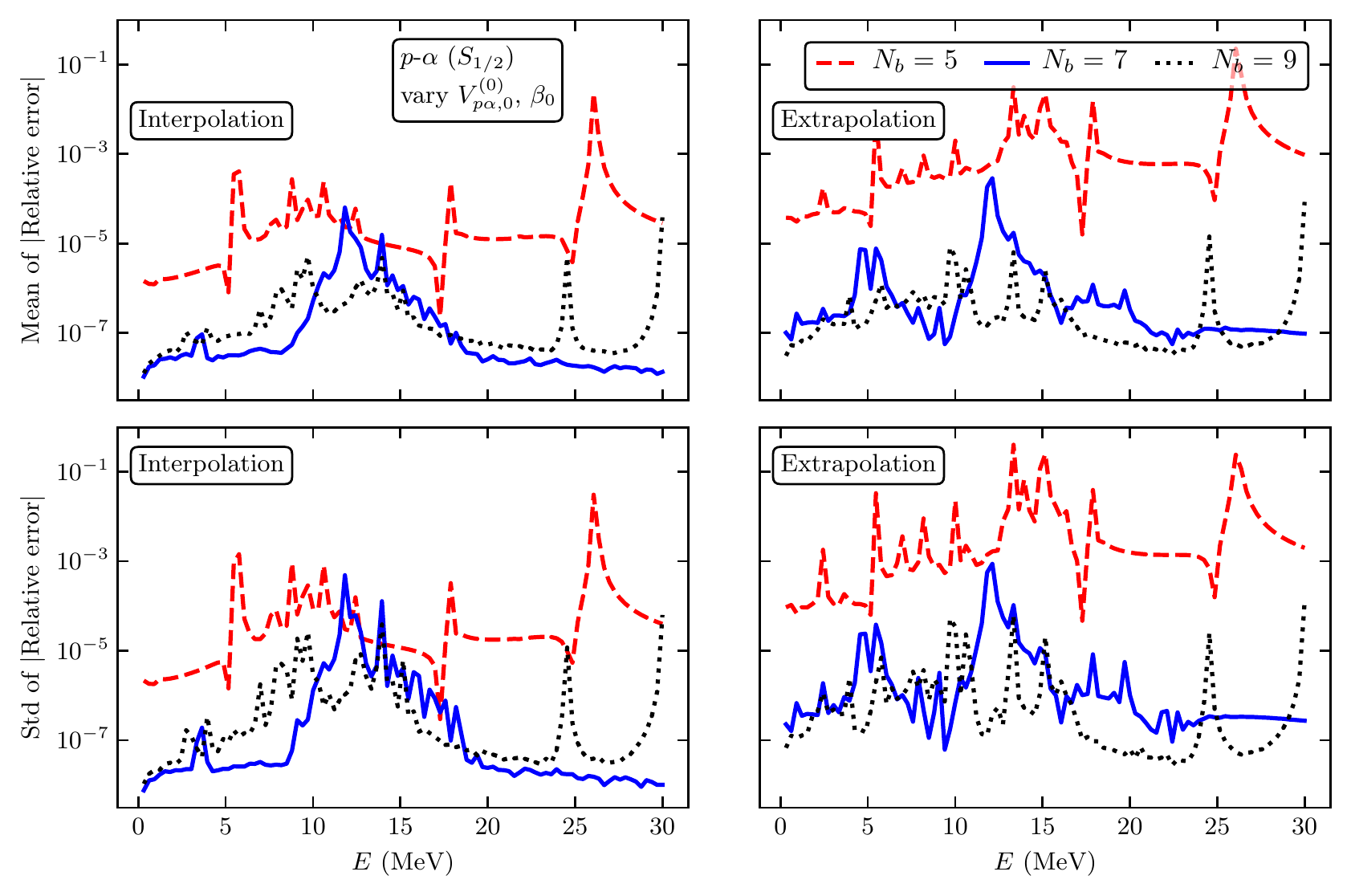}
  
    \caption{The relative errors for $\tan\delta(E)$ in the $p$--$\alpha$ $S_{1/2}$ channel in the two-dimensional space $\thetavec_i = \{V^{(0)}_{p\alpha,0},\beta_0\}$. The nugget is set to $10^{-8}$ here. }
  \label{fig:palpha2dimShalf}
\end{figure*}

\begin{figure*}[tbh]
  \includegraphics[width=0.85\textwidth]{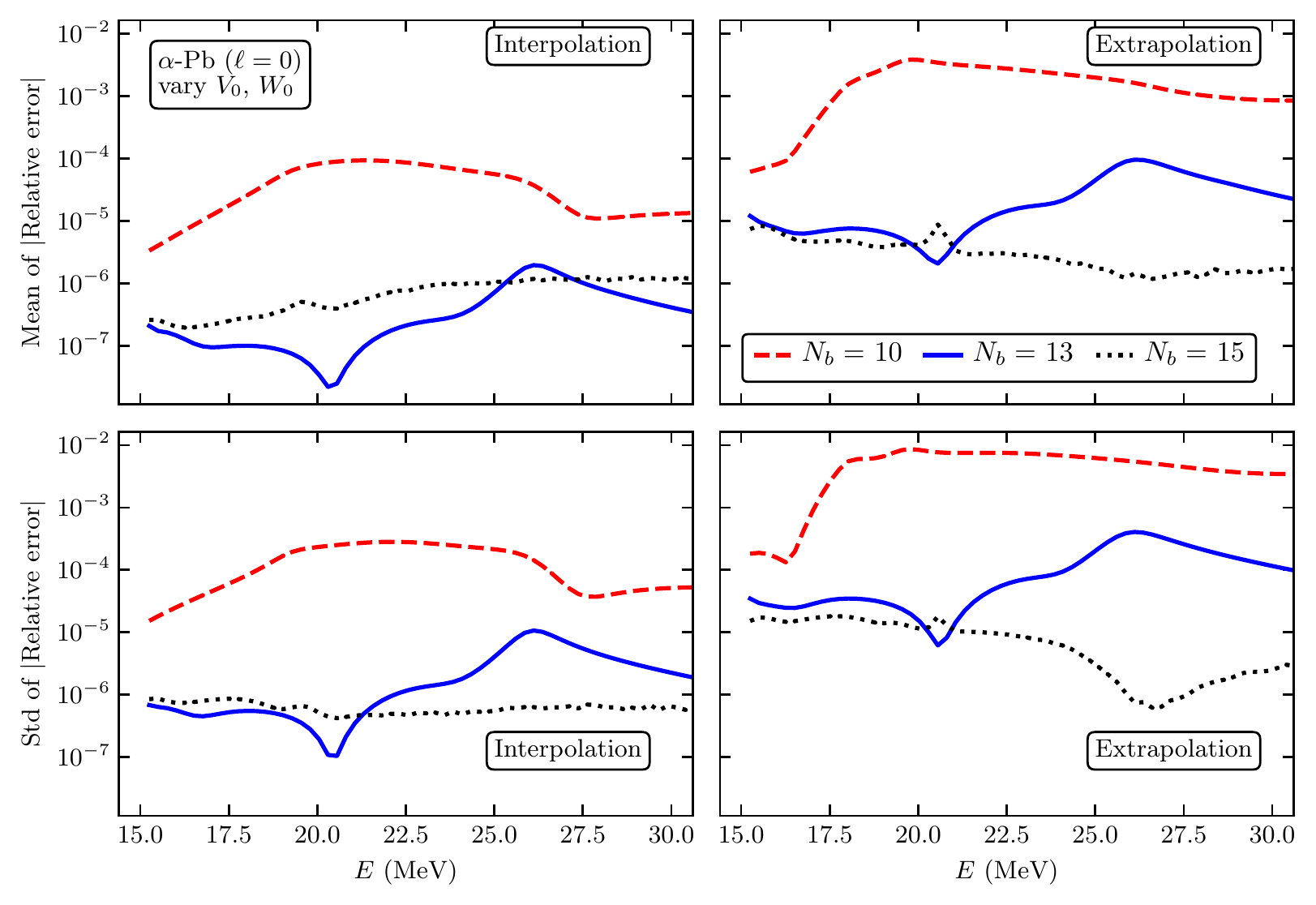}
  
    \caption{The relative errors for $\tan\delta(E)$ in the $\alpha$-$^{208}\mathrm{Pb}$ $\ell=0$ channel in the two-dimensional space $\thetavec_i = \{V_0,W_0\}$. Note that $\delta$ is complex. The nugget is set to $10^{-10}$ here.}
  \label{fig:alphaPb2diml0}
\end{figure*}

Here we provide additional results in the partial wave channels not already presented. The setups for these calculations are the same as for their counterparts in the main text.  Figures~\ref{fig:NN2dim3S1} and~\ref{fig:NN4dim3S1} provide results for NN scattering in the $^3S_1$ channel with the Minnesota potential by varying $\thetavec_i=\{V_{0R}, V_{0s}\}$ and  $\thetavec_i=\{V_{0R}, V_{0s}, \kappa_R, \kappa_R\}$, respectively. 
Figure~\ref{fig:palpha2dimShalf} shows the results for $p$-$\alpha$ scattering in the S-wave channel, in parallel to the P-wave results in the main text. Figure~\ref{fig:alphaPb2diml0} plots results for S-wave $\alpha$-$^{208}$Pb scattering. In these plots, both mean values and the standard deviations (std) of the relative errors (absolute values) are plotted against $E$. As mentioned in the main text and shown in these plots, the std values are similar to the mean values in general.

Please note we have publicized several jupyter notebooks, which can be used to reproduce all the results presented in this paper and this section. The notebooks together with necessary documentations can be accessed at~\url{https://github.com/buqeye/eigenvector-continuation}.

\end{document}